\documentclass{winnower}

\usepackage{epstopdf}
\usepackage{memhfixc}
\usepackage{graphicx}
\usepackage{lineno,hyperref}
\usepackage{amsmath}
\usepackage{latexsym}
\usepackage{amsfonts}
\usepackage{amssymb}
\usepackage{mathrsfs}
\usepackage[utf8]{inputenc}
\usepackage{bm}
\usepackage{graphicx}
\usepackage{hyperref}
\usepackage{subfigure}
\usepackage{pdfpages}
\newtheorem{theorem}{Theorem}[section]

\newtheorem{definition}{Definition}[section]
\newtheorem{lemma}[theorem]{Lemma}
\newtheorem{remark}{Remark}[section]
\newtheorem{corollary}[theorem]{Corollary}

\newcommand{\ba}{\begin{array}}
\newcommand{\ea}{\end{array}}

\newcommand{\bfl}{\begin{flushleft}}
\newcommand{\efl}{\end{flushleft}}
\newcommand{\bfr}{\begin{flushright}}
\newcommand{\efr}{\end{flushright}}
\newcommand{\bt}{\begin{theorem}}
\newcommand{\bd}{\begin{definition}}
\newcommand{\ed}{\end{definition}}
\newcommand{\et}{\end{theorem}}
\newcommand{\bl}{\begin{lemma}}
\newcommand{\el}{\end{lemma}}
\newcommand{\ee}{\end{exam}}
\newcommand{\bcor}{\begin{corollary}}
\newcommand{\ecor}{\end{corollary}}
\usepackage[utf8]{inputenc}
\hypersetup{colorlinks,linkcolor={red},citecolor={blue},urlcolor={red}} 
\usepackage{appendix}
\usepackage{fancyhdr}
 \usepackage{multirow}
\usepackage{tabularx}
\usepackage{makecell}
\usepackage{lscape}
\usepackage{float}
\usepackage{lipsum}
\usepackage{xcolor}
\usepackage{lipsum}
\makeatletter
\DeclareRobustCommand{\change}{%
  \@bsphack
  \leavevmode
  \color{red}%
  \@esphack
}
\DeclareRobustCommand{\stopchange}{%
  \@bsphack
  \normalcolor
  \@esphack
}
\makeatother

\begin{document}

\title{Resilience family of receiver operating characteristic curves}

\author{Ruhul Ali Khan}
\affil{Theoretical Statistics and Mathematics Unit, Indian Statistical Institute, Delhi Centre,  New Delhi 110016, India}

\date{}

\maketitle

\let\thefootnote\relax\footnotetext{E-mail Addresses: Ruhul Ali Khan (ruhulali.khan@gmail.com)}

\begin{abstract}
A new semiparametric model of the ROC curve based on the resilience family or proportional reversed hazard family is proposed which is an alternative to the existing models. The resulting ROC curve and its summary indices (such as area under the curve (AUC) and Youden index) have simple analytic forms. The partial likelihood method is applied to estimate the ROC curve. Moreover, the estimation methodologies of the resilience family of the ROC curve have been developed based on AUC estimators exploiting Mann-Whitney statistics and the Rojo approach. A simulation study has been carried out to assess the performance of all considered estimators. Real data  from the American National Health and Nutrition Examination Survey (NHANES)  has been analysed in detail based on the proposed model and the usual binormal model prevalent in the literature. Real data in the context of brain injury-related biomarkers is also analysed in order to compare our model with the Lehmann family of the ROC curves. Finally, we show that the proposed model may be applicable in the misspecification scenario through a  Ducheme muscular dystrophy data. \\

{\bf Keywords:} 
Semiparametric model, ROC curve, Area under the curve, Mann-Whitney statistics. \\

{\bf AMS Subject Classification:} Primary 92B15, Secondary 62P10.
\end{abstract}

%-------------------------------------------------%
\section{Introduction}\label{sec1}
%-------------------------------------------------%

In statistical decision theory, one of the important aspects is to construct a reasonable classification rule based on some characteristics. The receiver operating characteristic (ROC) curve represents the performance of a binary classifier. While the ROC curve originated in the analysis of radar signals (\cite{green1966signal}) and it has been consequently adopted and extended in diversified fields such as Diagnostic Medicine, psychology, Banking, Finance, biometrics, forensic sciences and many more (see \cite{pepe2003statistical}, \cite{krzanowski2009roc} and \cite{zhou2009statistical} for an overview).
In a biomedical environment, the true positive rate (TPR) is the probability that an individual from diseased group is correctly classified (sensitivity) and the false positive rate (FPR) is the probability that an individual is misclassified from healthy group (1 - specificity). The ROC curve is defined as a plot of TPRs against FPRs resulting from a continuous classification function for various threshold values. 

Suppose there are two populations- a ``positive" population $P$ and a ``negative" population $N$ together with a classification rule assumed to be continuous function $S(\bm{X})$ of the random vector $\bm{X}$ of variables measured on each individual. Let $t$ be the value of the threshold $T$ such that an individual is allocated to a population $P$ if the classification score $s(\bm{x})$ exceeds $t$ and otherwise to population $N$ where $\bm{x}$ is the observed value of $\bm{X}$. Note that $\text{FPR}(t)=p(S>t\vert N)=1-F_0(t)=\bar{F}_0(t)$ and $\text{TPR}(t)=p(S>t\vert P)=1-F(t)=\bar{F}(t)$ when $S$ possesses distribution function $F_0$ and $F$ for the population $N$ and $P$ respectively. Then the ROC curve graphs the trade-off between $\bar{F}(t)$ and $\bar{F}_0(t)$ or equivalently is a plot of 
\begin{equation}
\label{roccurve}
R(t) = 1-F(F_0^{-1}(1-t))    
\end{equation}
 against $t$ for $t\in [0, 1]$ and $F_0^{-1}(u)=\inf\left\lbrace x: F_0(x)\geq u \right\rbrace$. There are several indices in the literature for summarizing the information given by an ROC curve. The area under the ROC curve, denoted by AUC, is perhaps the most widely used summary index for the ROC curve and is defined by

\begin{equation}
\label{auc}
    \text{AUC}=\int_0^1 R(t)dt.
\end{equation}

Note that the plot of an ROC curve lies above the diagonal of the unit square and consequently the AUC takes value between 0.5 and 1, where 0.5 and 1 represent random classification and perfect classification respectively. The Youden index, denoted by $J$, depicts the maximum difference between the TPR and the FPR, i.e.,

\begin{equation}
\label{YI}
    J=\max_t\left\lbrace\text{TPR}(t)-\text{FPR}(t)\right\rbrace=\max_t\left\lbrace\text{sensitivity}(t)+\text{specificity}(t)-1\right\rbrace
\end{equation}
since TRP =  sensitivity and FPR = 1 - specificity. The threshold corresponding to the
Youden index $J$ is often taken to be the optimal classification threshold (\cite{youden1950index}, \cite{perkins2005youden}).

In the ROC curve analysis, the assumption of the binormal model is common where $F_0$ and $F$ are assumed to be normal distribution (see \cite{faraggi2002estimation} and the references therein) since the ROC curve remains the same if the classification scores undergo a strictly increasing transformation.  There are several parametric models for ROC curve in the literature such as bi-gamma, bi-beta, bi-logistic, bi-lognormal, bi-Rayleigh etc. (see \cite{gonccalves2014roc} for an overview). One of the simplest methods for estimating the ROC curve is the use of empirical ROC curve which can be obtained by plugging in empirical estimates of $F_0$ and $F$ into (\ref{roccurve}). The empirical ROC curve uniformly converges to the theoretical curve and preserves many interesting properties of the empirical distribution function (\cite{hsieh1996nonparametric}). \cite{zou1997smooth} used kernel density estimator for $F_0$ and $F$ to estimate the ROC curve which overcomes the lack of smoothness of the empirical estimator. Lehmann family of ROC was proposed by \cite{gonen2010lehmann} where they assumed that $\bar{F}=\bar{F}_0^\gamma$ for $\gamma\in (0, 1)$. In the last three decades, several researchers have proposed so many methods for estimating the ROC curve and its functionals. In this context, some of the references are \cite{zou2000two}, \cite{lloyd2002theory}, \cite{cai2002semiparametric}, \cite{qin2003using}, \cite{cai2004semi}, \cite{wan2007smooth}, \cite{davidov2012improving}, \cite{jokiel2013nonparametric},
 \citeauthor{jokiel2020estimation} (\citeyear{jokiel2020estimation}, \citeyear{jokiel2021minimum}). 

The organization of the paper is as follows. In section 2, a new semiparametric model of the ROC curve based on the resilience parameter family  or alternatively, a proportional reverse hazards family, with underlying distribution $F_0$, has been proposed which is an alternative to the existing models. The resulting ROC curve and its summary indices (such as AUC and Youden index) have simple analytic forms. Section 3 deals with estimation methodology of the ROC curve based on the proposed model. A brief discussion about the verification of the assumed model has been made at this juncture. The partial likelihood method is applied to estimate the ROC curve. Moreover, the estimation methodologies of the resilience family of the ROC curve have been developed based on AUC estimators expoiting Mann-Whitney statistics and the Rojo approach. Estimation procedure of the Youden index is also discussed. In Section 4, a simulation study has been carried out in order to assess the performance of all considered estimators. In section 5, a real data has been analysed based on the proposed model and existing models, and some remarks are made. Further, another real data in the context of brain injury-related biomarkers are also analysed in order to compare our model with the Lehmann family of the ROC curves. Finally, we showed that the proposed model may be applicable in the misspecification scenario through Ducheme muscular dystrophy data. Section 6 provides conclusions and some prospects of the present study.

\section{Proposed model}

One of the important properties of the ROC curve is that the curve remains unaltered if the classification scores undergo a strictly increasing transformation.  Based on the previous discussion, suppose $F_0$ and $F$ are associated with absolutely continuous random variables $X$ and $Y$ with respect to Lebesgue measure on the real line (after some suitable transformation, if necessary). If $F(t)$ is defined as 
\begin{equation}
\label{resi}
F(t)=\left[ F_0(t) \right]^\theta \quad \quad \theta>0,
\end{equation} then, $\theta$ is called a resilience parameter and $\left\lbrace F(\cdot \vert \theta), \theta>0\right\rbrace$ is a resilience
parameter family with underlying distribution $F_0$ (see \cite{marshall2007life}). It is interesting to note that the assumption (\ref{resi}) corresponds to the proportional reversed hazard rates (PRHR) assumption of the form $\frac{h(t)}{h_0(t)} = \theta$, where $h_0=f_0/F_0$ and $h=f/F$ are reversed hazard rate (RHR) of the random variables $X$ and $Y$ respectively (see \cite{von1936distribution}, \cite{barlow1963properties}, \cite{keilson1982uniform} and \cite{gupta2007proportional}). For a very small interval, the probability of failure in the interval given failure before the end of the interval is approximated by the product of the RHR and the length of the interval. \cite{kalbfleisch1989inference} showed that RHR is important in the estimation of the survival function under left censored data. \cite{block1998reversed} discussed some properties of the RHR function while \cite{nanda2001hazard} derived some interesting results which compare order statistics in the RHR orders. \cite{kundu2004characterizations} provided two simple characterizations of the PRHR class of distributions based on some conditional expectation and conditional variance. Discrete life distributions with decreasing RHR were studied by \cite{nanda2005discrete}. \cite{sengupta2010proportional} pointed out that PRHR is applicable where the PH model is inappropriate. \cite{wang2015interval} developed estimation methodology for the confidence intervals of the family of PRHR distributions based on lower record values. A proportional cause-specific reversed hazards model was introduced by \cite{sankaran2016proportional} and
\cite{fallah2021statistical} proposed statistical  inference  for  component  lifetime  distribution  from  coherent  system  lifetimes  under  a PRHR  model. \cite{baratnia2021reversed} proposed a new statistical method for one-way classification which is based on the RHR function of the response variable and \cite{khan2021some} studied the relationship between RHR and mean inactive time function. Ever since the inception of the RHR, due to its diverse applications, many researchers have contributed to it and a vast number of publications on this topic have appeared. Here, I have tried only to mention some significant works regarding RHR with a brief literature survey. For application in medical studies of RHR, one may refer to  \cite{kalbfleisch1991regression} or \cite{andersen2012statistical}. 

There exist many well known distributions which are belong to the resilience  parameter  family. Some of these are the Burr X distribution (\cite{burr1942cumulative}), Topp-Leone distribution (\cite{topp1955family}),  generalized exponential distribution (\cite{gupta1999theory}), Generalized Rayleigh distribution (\cite{kundu2005generalized}), generalized Gompertz distribution (\cite{el2013generalized}) and exponential-type distribution (\cite{lemonte2013new}).

\begin{figure}
    \centering
    \includegraphics[scale=0.7]{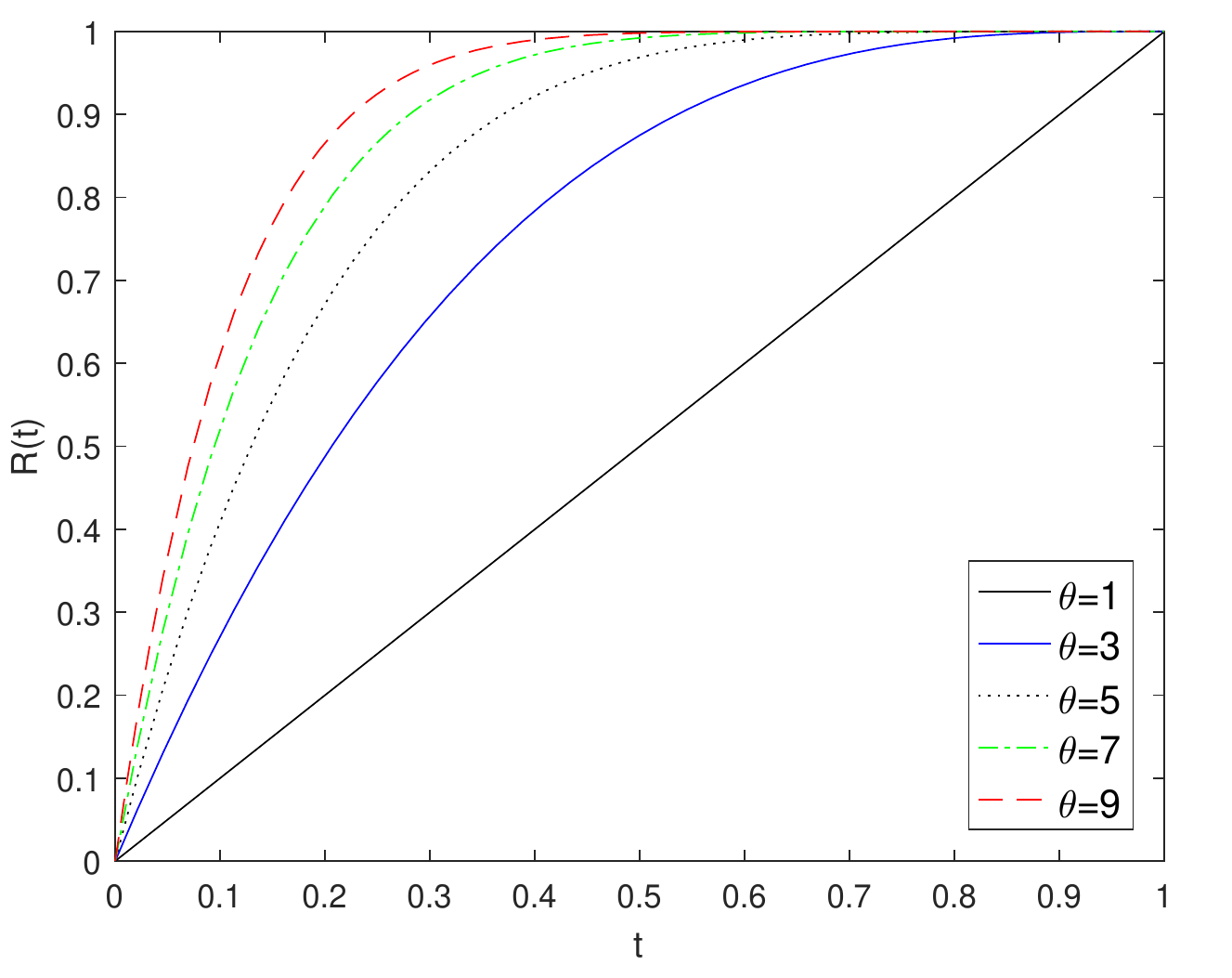}
    \caption{Resilience family of ROC curves with various value of $\theta$}
    \label{ROCtheta}
\end{figure}

From (\ref{roccurve}) and the assumption given in (\ref{resi}), the expression of the ROC curve reduces to the following very simple analytic form
\begin{equation}
\label{rocprhr}
R_\theta(t):= R(t)=1-\left(1-t\right)^\theta , \quad\quad t\in[0, 1],
\end{equation}
where $F_0$ and $F$ are any unknown distributions. For a sensible classifier it is important to consider that $\text{TPR}(t)>\text{FPR}(t)$. To ensure this fact, assume that $\theta>1$ for the proposed model. $\theta=1$ represents the chance diagonal, i.e., the diagonal of the unit square from $(0,0)$ to $(1, 1)$ which represents the random allocation. Thus the family
\begin{equation}
\mathcal{R}=\left\lbrace R_\theta : \theta\in (1, \infty) \right\rbrace    
\end{equation}
is called the {\it resilience family of ROC curves}.  Figure \ref{ROCtheta} shows the various ROC curves for some value of $\theta$. One can note that the large value of $\theta$ represents better classifier. 

Now several indices will be presented for summarizing the information given by the ROC curve in the context of proposed model. The area under the ROC curve given in (\ref{auc}) can also be represented in the following forms.

\begin{equation}
\label{aucform}
\text{AUC}= \int_0^1 R(t)dt = \int_0^1 \left[1-F(F_0^{-1}(1-t))\right] d t = \int_{-\infty}^{\infty} F_0(u) d F (u) = P(X<Y)
\end{equation}

Thus the area under the resilience family of ROC curves is given by 
\begin{equation}
\label{aucresi}  
\text{AUC}:=\tau= \frac{\theta}{1+\theta}
\end{equation}

From (\ref{YI}), after simplifying, the expression for the Youden index for the proposed model is given by,

\begin{equation}
\label{yiprhr}
J= \left(\frac{1}{\theta}\right)^{\frac{1}{\theta-1}}- \left(\frac{1}{\theta}\right)^{\frac{\theta}{\theta-1}}
\end{equation}
 and the corresponding optimum cut-off point is $t^*=F_0^{-1}\left[\left(\frac{1}{\theta}\right)^{\frac{1}{\theta-1}}\right]$.

The next section deals with estimation methodologies for $\theta$, $\tau$ and $J$.

\section{Estimation of the resilience family of ROC curves}

Suppose $\bm{X}_m = (X_1, X_2, \dots ,X_m)$ and $\bm{Y}_n = (Y_1, Y_2, \dots ,Y_n)$ are independent samples from $F_0$ and $F$ respectively where $F_0$ and $F$ are unknown. Let $F_{0m}$ and $F_n$ be the empirical distribution functions of $\bm{X}_m$ and $\bm{Y}_n$. Before going to estimation methodologies it is important to verify the assumption between $F_0$ and $F$ given in (\ref{resi}) based on the data. Now the estimation methodologies of $\theta$ will be presented in the sequel. 

\subsection{The maximum partial likelihood estimator}

Let $W_{(1)}\leq W_{(2)}< \dots \leq W_{(m+n)}$ be the ordered observation combining $X_i$, $i=1, 2, \dots, m$ and $Y_i$, $i=1, 2, \dots, n$. Suppose $x_p$ of the $X_i$'s and $y_p$ of the $Y_i$'s are less than or equal to $W_{(p)}$, $p=1, 2, \dots , m+n$. Now the proportional reversed hazard regression model will be introduced to study the relationship in presence of covariates. The proportional reversed hazard model for $i$-th individual is defined by
\begin{equation}
\label{prhrcov}
h_i(t\vert Z)=h_0(t)\exp{(\beta z_i)}
\end{equation}
where $e^{\beta}=\theta$ and $z_i\in\{0, 1\}$. Since the model considers the simple case of comparing positive group $P$ and negative group $N$, the covariate $z_i$ is either $1$ (if the individual is in the positive group) or $0$ (if the individual is in the negative group). Now note that 
\begin{equation}
\frac{h_0(t)\exp{(\beta z_i)}} {\sum_{C_i(t)}h_0(t)\exp{(\beta z_i)}}= \frac{\exp{(\beta z_i)}} {\sum_{C_i(t)}\exp{(\beta z_i)}}
\end{equation}
where $C_i(t)= \text{number of } W_i$'s less or equal to $t$. Thus the partial likelihood can be written as 
\begin{equation}
L(\theta, t)= \prod_{i=1}^{m+n} \frac{\exp{(\beta z_i)}} {\sum_{C_i(t)}\exp{(\beta z_i)}}.
\end{equation}
After some simplification log of the partial likelihood yields
\begin{equation}
l(\theta, t)= n \log{\theta} - \sum_{i=1}^{m+n} \log{\left(x_i+ y_i \theta\right)}
\end{equation}
Thus, as a result, the estimate of the parameter $\theta$ of the resilience ROC curve, denoted by $\hat{\theta}^{pl}$, can be obtained by solving the following equation numerically,
\begin{equation}
 \frac{n}{\theta}- \sum_{i=1}^{m+n}  \frac{y_i}{x_i+ y_i \theta}=0.
\end{equation}
For an example in R, one may refer to `nleqslv' package (\cite{nleq}) to solve the equation since the solution of the above equation does not have an explicit form. 

Now We will proceed to test $H_0: \theta=1$ against $\theta\neq 1$ and construct $100(1-\alpha)$\% confidence interval based on the Wald's test. The observed information $I(\hat{\theta}^{pl})$ yields from
\begin{equation}
 I(\theta)=-\frac{\partial^2 l}{\partial \theta^2}= \frac{n}{\theta^2}-  \sum_{i=1}^{m+n} \frac{y_i^2}{\left(x_i+ y_i \theta\right)^2}   
\end{equation}
by plugging in $\hat{\theta}^{pl}$. The variance of $\hat{\theta}^{pl}$ is approximately $1/I(\hat{\theta}^{pl})$ and the standard error $\text{S.E.}(\hat{\theta}^{pl})=1/\sqrt{I(\hat{\theta}^{pl})}$. Now, we use $\text{S.E.}(\hat{\theta}^{pl})$ to construct a normalized test statistic $Z_w=\frac{\hat{\theta}^{pl}-1}{\text{S.E.}(\hat{\theta}^{pl})}$ and reject $H_0$ if $\vert Z_w \vert >Z_{\alpha/2}$ where $Z_{\alpha}$ is the upper $\alpha$-th quantile of the standard normal distribution. Thus the $100(1-\alpha)$\% confidence interval is given by $\left(\hat{\theta}^{pl} - Z_{\alpha/2} \cdot \text{S.E.}(\hat{\theta}^{pl}), \hat{\theta}^{pl} + Z_{\alpha/2} \cdot \text{S.E.}(\hat{\theta}^{pl}) \right)$.

\begin{remark}
The estimator of AUC, $\hat{\tau}^{pl}=\frac{\hat{\theta}^{pl}}{1+\hat{\theta}^{pl}}$ and using delta method we obtain the $100(1-\alpha)$\% confidence interval of $\tau$ which is given by $\left(\hat{\tau}^{pl} - Z_{\alpha/2} \cdot \frac{\text{S.E.}(\hat{\theta}^{pl})}{1+\hat{\theta}^{pl}}, \hat{\tau}^{pl} + Z_{\alpha/2} \cdot \frac{\text{S.E.}(\hat{\theta}^{pl})}{\hat{\theta}^{pl}} \right)$.
\end{remark}

\begin{remark}
From (\ref{yiprhr}), the estimator for the Youden index is given by
\begin{equation}
\label{estyi}
\hat{J}^{pl}= \left(\frac{1}{\hat{\theta}^{pl}}\right)^{\frac{1}{\hat{\theta}^{pl}-1}}- \left(\frac{1}{\hat{\theta}^{pl}}\right)^{\frac{\hat{\theta}^{pl}}{\hat{\theta}^{pl}-1}}.
\end{equation}
 %and the corresponding estimate of the optimum cut-off point is $F_{0m}^{-1}\left[\left(\frac{1}{\hat{\theta}^{pl}}\right)^{\frac{1}{\hat{\theta}^{pl}-1}}\right]$. 
 One can also find confidence interval for the Youden index using delta method.
\end{remark}

\begin{remark}
It is possible that the calculated value of $\hat{\theta}^{pl}$ may be found to be less than $1$ when $\theta$ is close to $1$ and the sample size is small. In this situation one can consider $\min\{1, \hat{\theta}^{pl}\}$ as an estimator to obtain smaller bias and variance.
\end{remark}

\subsection{Estimation based on the Mann–Whitney statistics}
The expression given in (\ref{aucform}) suggests to propose an estimator of the AUC ($\tau$) using Mann–Whitney statistics and derive its asymptotic distribution exploiting U-statistics theory.  Let the kernel be $h(x, y) = I(x < y)$, with expectation $\tau = P(X <Y)$ where  $I(x < y)=1$ if $x < y$, $0$ otherwise. Then, the corresponding U-statistic is
\begin{equation}
\label{Uroc}
\hat{\tau}^{MW}=\frac{1}{mn}\sum_{i=1}^m \sum_{j=1}^n h(X_i, Y_i).
\end{equation}
Assume that there is no ties. Then $\hat{\tau}^{MW}$ is the Mann–Whitney statistics which is an unbiased nonparametric estimator of $\tau$.
Now one can obtain the following theorem by considering the projection of $\hat{\tau}^{MW}$ viewed as a two-sample U-statistic (\cite{serfling1980}[page 193]).
\begin{theorem}
\label{serf}
Let $m, n \rightarrow \infty$ in such a way that $m/(m + n) \rightarrow p$, $0 < p < 1$. Suppose that $\sigma_{10}^2=P\left( X<Y, X^\prime <Y\right)-P(X<Y)^2<\infty$ and $\sigma_{01}^2=P\left( X<Y, X<Y^\prime\right)-P(X<Y)^2<\infty$. Then
$$\sqrt{m+n}\left[\hat{\tau}^{MW} -\tau \right] \xrightarrow[]{d} \mathcal{N}(0, \sigma^2) \quad \text{as } \min{(m, n)} \rightarrow \infty,$$
where $\sigma^2=\frac{1}{p}\sigma_{10}^2+\frac{1}{1-p}\sigma_{01}^2$.
\end{theorem}
\begin{remark}
 It is worthwhile noting that the above theorem can also be obtained by applying Theorem 2.3 of \cite{hsieh1996nonparametric}.
\end{remark}
In the context of the proposed model, the simplified expressions of $\sigma_{10}^2$ and $\sigma_{01}^2$ are: 

$$\sigma_{10}^2=\frac{\theta}{(2+\theta)(1+\theta)^2} \text{ and } \sigma_{01}^2=\frac{\theta^2}{(1+2\theta)(1+\theta)^2}.$$
Thus Theorem \ref{serf} reduces to the following theorem.
\begin{theorem}
\label{aucman}
Let $m, n \rightarrow \infty$ in such a way that $m/(m + n) \rightarrow p$, $0 < p < 1$. Then for $\sigma^2>0$,
$$\sqrt{m+n}\left[\hat{\tau}^{MW} -\tau \right] \xrightarrow[]{d} \mathcal{N}(0, \sigma^2(\tau)) \quad \text{as } \min{(m, n)} \rightarrow \infty,$$
where $\sigma^2(\tau)=\frac{1}{p}\frac{\tau (1-\tau)^2}{(2-\tau)}+\frac{1}{1-p}\frac{\tau^2(1-\tau)}{(1+\tau)^2}$.
\end{theorem}

\begin{remark}
\label{citaumw}
The $100(1-\alpha)$\% confidence interval of $\tau$ is given by 
$$\left[\hat{\tau}^{MW}- \frac{Z_{\alpha/2}}{\sqrt{m+n}} \cdot \sigma^2(\hat{\tau}^{MW}), \hat{\tau}^{MW}+ \frac{Z_{\alpha/2}}{\sqrt{m+n}} \cdot \sigma^2(\hat{\tau}^{MW})\right]$$
since $\hat{\tau}^{MW}$ is a consistent estimator of $\tau$.
\end{remark}

From relation (\ref{aucresi}), the estimator of $\theta$ will be proposed by plug-in $\hat{\tau}^{MW}$. Thus the estimator of $\theta$ based on  Mann–Whitney statistics is given by

$$\hat{\theta}^{MW}=\frac{1}{1-{mn}/{\sum_{i=1}^m \sum_{j=1}^n h(X_i, Y_i)}}.$$

\begin{remark}
An application of Jensen inequality and the fact that $g(x)=\frac{x}{1-x}$ is a convex function on $(0,1)$ yields $E(\hat{\theta}^{MW})\geq \theta$. Thus $\hat{\theta}^{MW}$ is a biased estimator for $\theta$ but $\hat{\tau}^{MW}$ is an unbiased estimator for $\tau$.  
\end{remark}
Now an application of delta method yields the following theorem which establishes the asymptotic normality of $\hat{\theta}^{MW}$. 

\begin{theorem}
\label{rojotheta}
Under the assumption of Theorem \ref{aucman},
$$\sqrt{m+n}\left[\hat{\theta}^{MW} -\theta \right] \xrightarrow[]{d} \mathcal{N}(0, \sigma^2(\theta)) \quad \text{as } \min{(m, n)} \rightarrow \infty,$$
where $\sigma^2(\theta)=\frac{1}{p}\frac{\theta (1+\theta)^2}{(2+\theta)}+\frac{1}{1-p}\frac{\theta^2 (1+\theta)^2}{(1+2\theta)}$.
\end{theorem}

\begin{remark}
Similarly as in Remark \ref{citaumw}, the $100(1-\alpha)$\% confidence interval of $\theta$ is given by 
$$\left[\hat{\theta}^{MW}- \frac{Z_{\alpha/2}}{\sqrt{m+n}} \cdot \sigma^2(\hat{\theta}^{MW}), \hat{\theta}^{MW}+ \frac{Z_{\alpha/2}}{\sqrt{m+n}} \cdot \sigma^2(\hat{\theta}^{MW})\right].$$
\end{remark}

\begin{remark}
In this context, the proposed estimator for the Youden index $\hat{J}^{MW}$ can be obtained by replacing $\hat{\theta}^{pl}$ by $\hat{\theta}^{MW}$ in the right hand side expression of (\ref{estyi}). % and the corresponding estimate of the optimum cut-off point is $F_{0m}^{-1}\left[\left(\frac{1}{\hat{\theta}^{MW}}\right)^{\frac{1}{\hat{\theta}^{MW}-1}}\right]$. 
Moreover, an application of delta method can also produce asymptotic normality and confidence interval for this estimator. For the sake of brevity, the presentation of the asymptotic normality theorem has been avoided. 
\end{remark}

\subsection{Estimation based on Rojo approach}

The resilience family of ROC curve satisfies $X\leq_{st} Y$, i.e., $F_0\geq F$. This motivates us to estimate $F_0$ and $F$ by imposing order restricted condition. In this context, \cite{lo1987estimation} proposed the estimators of $F_0$ and $F$ as $F_{0m}^L=\max\{F_{0m}, F_n\}$ and $F_n^L=\min\{F_{0m}, F_n\}$ respectively since the the estimators satisfy the constraint of interest. But \cite{rojo2004estimation} pointed out some drawbacks of the estimators given in \cite{lo1987estimation} and proposed new estimators. Thus, the estimators of $F_0$ and $F$ based on Rojo approach will be used in this subsection. 

From (\ref{aucform}), it is easy to obtain $\tau=\int_0^1 F_0\left(F^{-1}(t) \right) d t$ and the proposed estimator for $\tau$ is given by,
\begin{equation}
\label{estrojo}
    \hat{\tau}^{R}=\int_0^1 F_{0mn}\left(F_{mn}^{-1}(t) \right) d t = \int_{\infty}^{\infty} F_{0mn}(t) d F_{mn} (t)
\end{equation}

where 
$$F_{0mn}(t)=\max\{F_{0m}(t), P_{mn}(t) \},$$
$$F_{mn}(t)=\min\{F_{n}(t), P_{mn}(t) \}$$
and
$P_{mn}$ is the empirical cumulative distribution function based on the
combined samples, i.e.,
$$P_{mn}(t)=\frac{m}{m+n}F_{0m}(t)+ \frac{n}{m+n}F_{n}(t).$$

Now the following theorem can be obtained using similar argument given in the proof of Theorem 2 of \cite{jokiel2020estimation}.
\begin{theorem}
\label{equirojo}
Under the assumption of Theorem \ref{aucman},
$$\sqrt{m+n}\left[\hat{\tau}^{R} -\tau \right] \xrightarrow[]{d} \mathcal{N}(0, \sigma^2(\tau)) \quad \text{as } \min{(m, n)} \rightarrow \infty,$$
where $\sigma^2(\tau)=\frac{1}{p}\frac{\tau (1-\tau)^2}{(2-\tau)}+\frac{1}{1-p}\frac{\tau^2(1-\tau)}{(1+\tau)^2}$.

%The estimator $\hat{\tau}^{R}$ given in (\ref{Uroc}) is asymptotically equivalent to the estimator $\hat{\tau}^{MW}$ given in (\ref{estrojo}).
\end{theorem}

From relation (\ref{aucresi}), the estimator of $\theta$ based on  Rojo approach is given by $\hat{\theta}^{R}=\frac{\hat{\tau}^{R}}{1-\hat{\tau}^{R}}$ and an application of delta method yields the following theorem.

\begin{theorem}
Under the assumption of Theorem \ref{rojotheta},
$$\sqrt{m+n}\left[\hat{\theta}^{R} -\theta \right] \xrightarrow[]{d} \mathcal{N}(0, \sigma^2(\theta)) \quad \text{as } \min{(m, n)} \rightarrow \infty,$$
where $\sigma^2(\theta)=\frac{1}{p}\frac{\theta (1+\theta)^2}{(2+\theta)}+\frac{1}{1-p}\frac{\theta^2 (1+\theta)^2}{(1+2\theta)}$.
\end{theorem}

\begin{table}[H]
\label{simutab}
\caption{Estimated values for different choices of $\theta$ based on Pl method, Mann-Whitney statistics and Rojo approach. Results are reported based on 10000 replications for each model and method.}
\centering
\begin{tabular}{cllllccccc} 
\hline
\begin{tabular}[c]{@{}c@{}}\\$\theta$\end{tabular} & \multicolumn{2}{c}{$(m, n)$}                   & \multicolumn{1}{c}{Method} & Avg($\hat{\theta}$) & SD($\hat{\theta}$)   & RMSE($\hat{\theta}$) & \begin{tabular}[c]{@{}c@{}}Coverage \\probability\end{tabular} & Avg($\hat{\tau}$)    & \multicolumn{1}{l}{Avg($\hat{J})$}  \\ 
\hline
\multirow{11}{*}{2}                                & \multicolumn{2}{c}{\multirow{3}{*}{(60, 60)}}  & Pl                         & 2.0581              & 0.4193               & 0.4232               & 0.9476                                                         & 0.6730               & 0.2599                              \\
                                                   & \multicolumn{2}{c}{}                           & MW                         & 2.0712              & 0.4835               & 0.4887               & 0.9434                                                         & 0.6744               & 0.2621                              \\
                                                   & \multicolumn{2}{c}{}                           & Rojo                       & 2.0759              & 0.4795               & 0.4855               & 0.9482                                                         & 0.6749               & 0.2629                              \\
                                                   &  &                                             &                            &                     & \multicolumn{1}{l}{} & \multicolumn{1}{l}{} & \multicolumn{1}{l}{}                                           & \multicolumn{1}{l}{} & \multicolumn{1}{l}{}                \\
                                                   & \multicolumn{2}{c}{\multirow{3}{*}{(60, 80)}}  & Pl                         & 2.0450              & 0.3872               & 0.3898               & 0.9433                                                         & 0.6716               & 0.2577                              \\
                                                   & \multicolumn{2}{c}{}                           & MW                         & 2.0627              & 0.4544               & 0.4587               & 0.9347                                                         & 0.6735               & 0.2607                              \\
                                                   & \multicolumn{2}{c}{}                           & Rojo                       & 2.0663              & 0.4516               & 0.4564               & 0.9389                                                         & 0.6739               & 0.2613                              \\
                                                   &  &                                             &                            &                     & \multicolumn{1}{l}{} & \multicolumn{1}{l}{} & \multicolumn{1}{l}{}                                           & \multicolumn{1}{l}{} & \multicolumn{1}{l}{}                \\
                                                   & \multicolumn{2}{c}{\multirow{3}{*}{(60, 100)}} & Pl                         & 2.0390              & 0.3696               & 0.3716               & 0.9463                                                         & 0.6710               & 0.2567                              \\
                                                   & \multicolumn{2}{c}{}                           & MW                         & 2.0575              & 0.4340               & 0.4378               & 0.9355                                                         & 0.6730               & 0.2598                              \\
                                                   & \multicolumn{2}{c}{}                           & Rojo                       & 2.0607              & 0.4316               & 0.4358               & 0.9386                                                         & 0.6733               & 0.2603                              \\
\multicolumn{1}{l}{}                               &  &                                             &                            &                     & \multicolumn{1}{l}{} & \multicolumn{1}{l}{} & \multicolumn{1}{l}{}                                           & \multicolumn{1}{l}{} & \multicolumn{1}{l}{}                \\
\multirow{11}{*}{4}                                & \multicolumn{2}{c}{\multirow{3}{*}{(60, 60)}}  & Pl                         & 4.1792              & 0.9902               & 1.0062               & 0.9467                                                         & 0.8069               & 0.4851                              \\
                                                   & \multicolumn{2}{c}{}                           & MW                         & 4.2285              & 1.1784               & 1.2003               & 0.9385                                                         & 0.8087               & 0.4885                              \\
                                                   & \multicolumn{2}{c}{}                           & Rojo                       & 4.2300              & 1.1777               & 1.1999               & 0.9392                                                         & 0.8088               & 0.4886                              \\
                                                   &  &                                             &                            &                     & \multicolumn{1}{l}{} & \multicolumn{1}{l}{} & \multicolumn{1}{l}{}                                           & \multicolumn{1}{l}{} & \multicolumn{1}{l}{}                \\
                                                   & \multicolumn{2}{c}{\multirow{3}{*}{(60, 80)}}  & Pl                         & 4.1394              & 0.9062               & 0.9168               & 0.9479                                                         & 0.8054               & 0.4824                              \\
                                                   & \multicolumn{2}{c}{}                           & MW                         & 4.2067              & 1.1073               & 1.1263               & 0.9318                                                         & 0.8079               & 0.4870                              \\
                                                   & \multicolumn{2}{c}{}                           & Rojo                       & 4.2078              & 1.1068               & 1.1261               & 0.9322                                                         & 0.8080               & 0.4871                              \\
                                                   &  &                                             &                            &                     & \multicolumn{1}{l}{} & \multicolumn{1}{l}{} & \multicolumn{1}{l}{}                                           & \multicolumn{1}{l}{} & \multicolumn{1}{l}{}                \\
                                                   & \multicolumn{2}{c}{\multirow{3}{*}{(60, 100)}} & Pl                         & 4.1251              & 0.8723               & 0.8812               & 0.9498                                                         & 0.8049               & 0.4814                              \\
                                                   & \multicolumn{2}{c}{}                           & MW                         & 4.1958              & 1.0666               & 1.0843               & 0.9253                                                         & 0.8075               & 0.4863                              \\
                                                   & \multicolumn{2}{c}{}                           & Rojo                       & 4.1969              & 1.0661               & 1.0841               & 0.9253                                                         & 0.8076               & 0.4863                              \\
\multicolumn{1}{l}{}                               &  &                                             &                            &                     & \multicolumn{1}{l}{} & \multicolumn{1}{l}{} & \multicolumn{1}{l}{}                                           & \multicolumn{1}{l}{} & \multicolumn{1}{l}{}                \\
\multirow{11}{*}{6}                                & \multicolumn{2}{c}{\multirow{3}{*}{(60, 60)}}  & Pl                         & 6.3505              & 1.7011               & 1.7367               & 0.9479                                                         & 0.8639               & 0.5964                              \\
                                                   & \multicolumn{2}{c}{}                           & MW                         & 6.4738              & 2.1064               & 2.1590               & 0.9358                                                         & 0.8662               & 0.6011                              \\
                                                   & \multicolumn{2}{c}{}                           & Rojo                       & 6.4749              & 2.1061               & 2.1588               & 0.9360                                                         & 0.8662               & 0.6011                              \\
                                                   &  &                                             &                            &                     & \multicolumn{1}{l}{} & \multicolumn{1}{l}{} & \multicolumn{1}{l}{}                                           & \multicolumn{1}{l}{} & \multicolumn{1}{l}{}                \\
                                                   & \multicolumn{2}{c}{\multirow{3}{*}{(60, 80)}}  & Pl                         & 6.2840              & 1.5540               & 1.5796               & 0.9478                                                         & 0.8627               & 0.5938                              \\
                                                   & \multicolumn{2}{c}{}                           & MW                         & 6.4387              & 1.9887               & 2.0364               & 0.9312                                                         & 0.8656               & 0.5998                              \\
                                                   & \multicolumn{2}{c}{}                           & Rojo                       & 6.4398              & 1.9884               & 2.0363               & 0.9315                                                         & 0.8656               & 0.5998                              \\
                                                   &  &                                             &                            &                     & \multicolumn{1}{l}{} & \multicolumn{1}{l}{} & \multicolumn{1}{l}{}                                           & \multicolumn{1}{l}{} & \multicolumn{1}{l}{}                \\
                                                   & \multicolumn{2}{c}{\multirow{3}{*}{(60, 100)}} & Pl                         & 6.2560              & 1.4921               & 1.5138               & 0.9491                                                         & 0.8622               & 0.5927                              \\
                                                   & \multicolumn{2}{c}{}                           & MW                         & 6.4169              & 1.9055               & 1.9505               & 0.9229                                                         & 0.8652               & 0.5989                              \\
                                                   & \multicolumn{2}{c}{}                           & Rojo                       & 6.4178              & 1.9053               & 1.9504               & 0.9230                                                         & 0.8652               & 0.5990                              \\
\hline
\end{tabular}
\end{table}

\begin{remark}
\label{citaur}
The $100(1-\alpha)$\% confidence interval of $\tau$ and $\theta$ are given by 
$$\left[\hat{\tau}^{R}- \frac{Z_{\alpha/2}}{\sqrt{m+n}} \cdot \sigma^2(\hat{\tau}^{R}), \hat{\tau}^{R}+ \frac{Z_{\alpha/2}}{\sqrt{m+n}} \cdot \sigma^2(\hat{\tau}^{R})\right]$$
and $$\left[\hat{\theta}^{R}- \frac{Z_{\alpha/2}}{\sqrt{m+n}} \cdot \sigma^2(\hat{\theta}^{R}), \hat{\theta}^{R}+ \frac{Z_{\alpha/2}}{\sqrt{m+n}} \cdot \sigma^2(\hat{\theta}^{R})\right]$$
respectively.
\end{remark}

%The following remarks are consequence of Theorem \ref{equirojo}.

\begin{remark}
In this context also, the proposed estimator for the Youden index $\hat{J}^{R}$ can be obtained by replacing $\hat{\theta}^{pl}$ by $\hat{\theta}^{R}$ in the right hand side expression of (\ref{estyi}).
%and the corresponding estimate of the optimum cut-off point is $F_{0m}^{-1}\left[\left(\frac{1}{\hat{\theta}^{R}}\right)^{\frac{1}{\hat{\theta}^{R}-1}}\right]$. %Moreover, an application of delta method can also produce asymptotic normality and confidence interval for this estimator. For the shake of brevity, we are not mentioning the asymptotic normality theorem.
\end{remark}

%\pagebreak

\section{Simulation study}

In this section, the performance of the three proposed estimators will be investigated by means of a simulation study. For this purpose, we choose the well known generalized exponential distribution (GED) which was introduced by \cite{gupta1999theory}. The cumulative distribution function (cdf) of GED is given by 

\begin{equation}
F(t)=\left(1-e^{-\lambda t}\right)^\theta, \quad t>0,
\end{equation}
where $\lambda>0$ and $\theta>0$ are the scale and resilience parameters respectively and are denoted by GED$(\lambda,\theta)$.  Note that GED belongs to the resilience  parameter  family  with underlying distribution $F_0(t)=1-e^{-\lambda t}$, i.e., the exponential distribution. The values of $\theta= 2, 4, 6$ and $\lambda=1$ were considered for the simulation study. For each value of the parameter $\theta$, we generate random observations of size $(m, n)=(60, 60)$,  $(60, 80)$ and $(60, 100)$ from $\left(\text{GED}(1, 1\right), \text{GED}(1,\theta)$. The proposed estimators were calculated for each pair of $(m,n)$. All the simulation study is performed using R (\cite{asdde}) on PC platform and the results are reported based on $10,000$ replications. The calculated values of $\tau$ are 0.6667,  0.8 and  0.8571  and $J$ are  0.25,  0.4725 and 0.5824 corresponding to $\theta$= 2, 4 and 6.

In Table \ref{simutab}, performance of different methods is investigated in terms of average value, standard deviation (SD), root mean square error (RMSE) and coverage probability of $\theta$. The reported coverage probabilities are based on significance level $\alpha=0.05$. The estimated values of AUC and Youden index are also presented. For GED model, observe that the Partial likelihood method works slightly better than other methods and all the estimators overestimate the resilience parameter $\theta$. The behaviour of the estimators based on Mann-Whitney statistics and Rojo approach are almost same in this case. Another interesting observation is that all the estimators perform well for small values of $\theta$.

\section{Analysis of real data}
Before going to analyse real data sets, it is necessary to verify the PRHR assumption for the samples. For this purpose it is necessary to develop a test for the verification of the PRHR assumption by the samples. In this paper, I would like to present a graphical representation for checking PRHR assumption. At first the verification of the usual stochastic dominance between $F_{0m}$ and $F_n$ via empirical plots should be made. After observing  $F_{0m}\geq F_n$, one should proceed to observe the plots of $\log{\left(-\log{\left(F_n(t)\right)}\right)}$ and $\log{\left(-\log{\left(F_{0m}(t)\right)}\right)}$, denoted by log-log plot. If the plot shows almost same difference between $\log{\left(-\log{\left(F_n(t)\right)}\right)}$ and $\log{\left(-\log{\left(F_{0m}(t)\right)}\right)}$ then the data may support the assumption given in (\ref{resi}) graphically. In the next subsections, two real data sets have been analysed in the sequel.  

\subsection{Diabetes data from NHANES}
In this section, data from the American National Health and Nutrition Examination Survey (NHANES) in 2009-2010 has been analysed (see \url{http://www.cdc.gov/nchs/data/series/sr_02/sr02_162.pdf}). The data is available in the `NHANES' package of R statistical software (\cite{NH}). The NHANES data includes $75$ variables with $5,000$ individuals of all ages.  Here, `Age' as a demographic variable, `Body Mass Index (BMI)' as a physical measurement, `Total Cholesterol (TotChol)' and `Diabetes'  as health variables have been considered in order to apply logistic regression as a binary classifier where  `Diabetes' is the qualitative response and the predictors are `Age', `BMI' and  `TotChol'. After removing `NA(s)' in  `Age', `BMI', `TotChol'  and `Diabetes', the data contains information of $4209$ individuals. Here the data arising from log odds has been used as a classification score by considering diabetes as the `positive' population (diseased, denoted by $1$) and those who did not suffer due to diabetes as `negative' population (non-diseased, denoted by $0$). 

\begin{figure}[ht]
  \begin{minipage}{0.45\textwidth}
    \includegraphics[scale=.45]{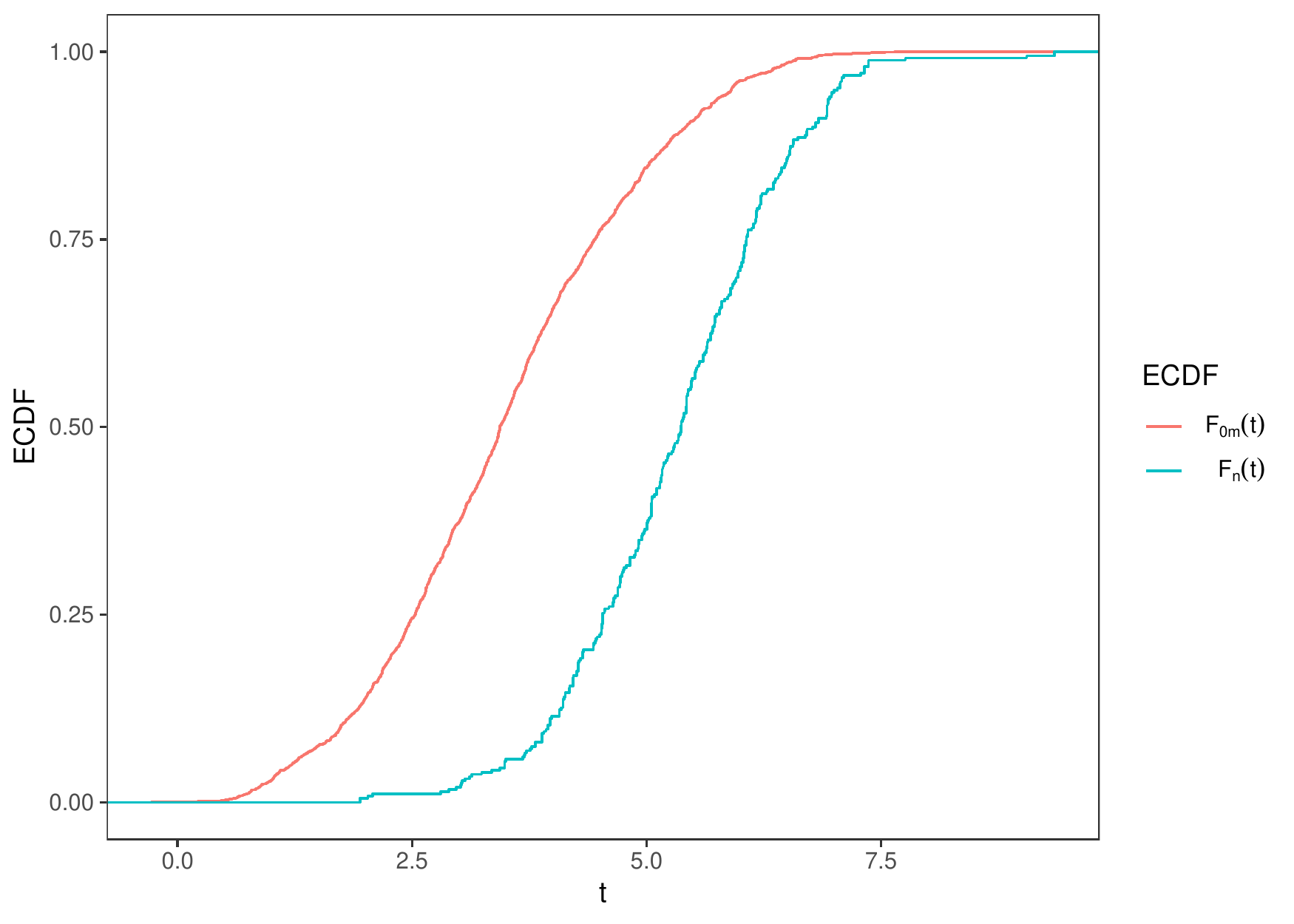}
    \label{fig1}
  \end{minipage}
  \hspace{.5cm}
  \begin{minipage}{0.45\textwidth}
    \includegraphics[scale=.5]{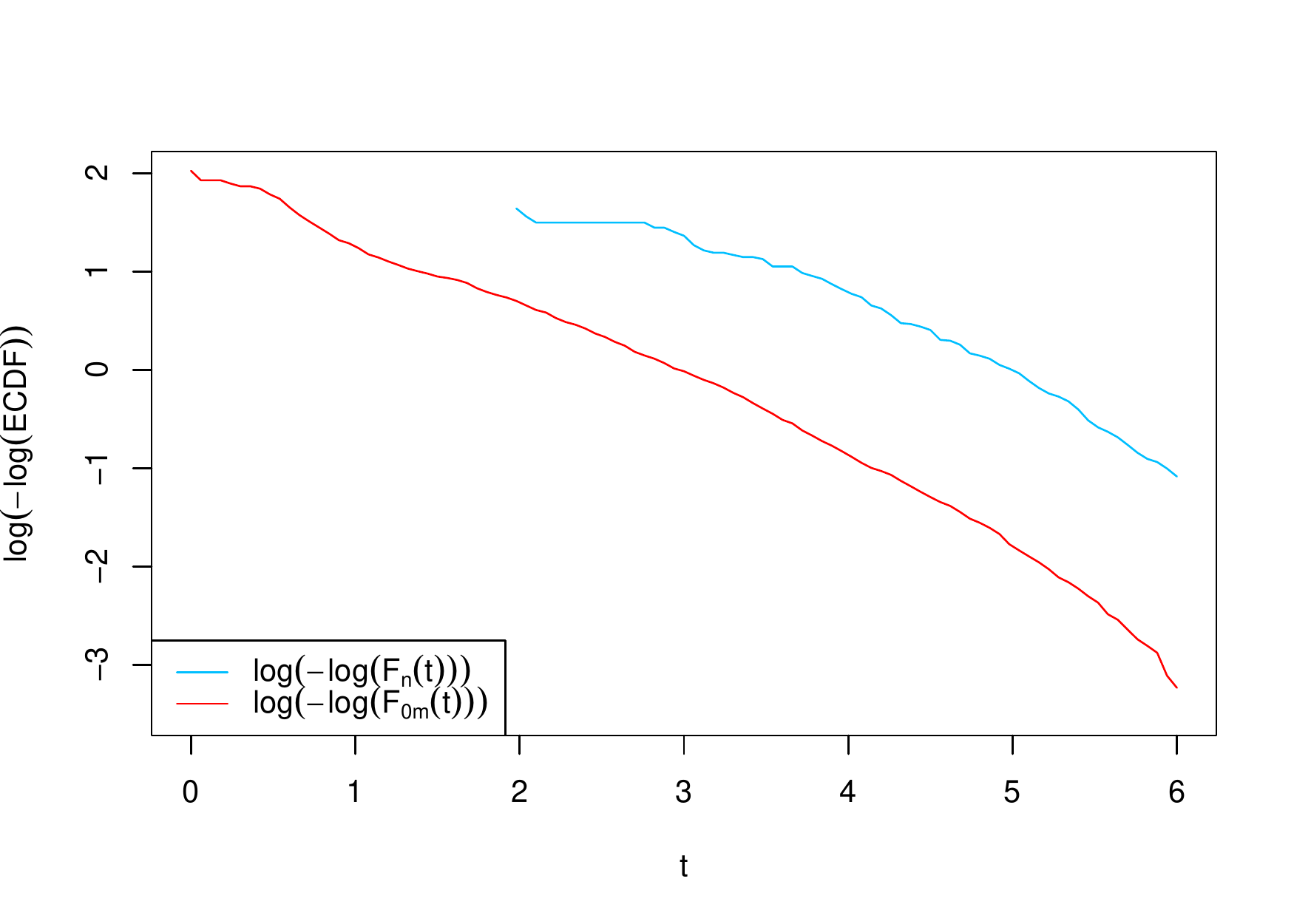}
    \label{fig2}
  \end{minipage}
      \caption{Empirical distribution functions and log-log plot}
      \label{ecdfloglog}
\end{figure}

One of the interesting properties of the ROC curve is that the ROC curve remains same if the classification scores undergo a increasing transformation. Thus, in order to transform all the values of the classification score in the positive domain, $\Phi(S)=6.85+S$ has been chosen as a transformation. The use of this transformation has been considered for estimation of the resilience ROC curve only. Now in the above setup, a single binary covariate $z_i \in \{0, 1\}$, the empirical distribution functions and the empirical $\log-\log$ functions are plotted in Figure \ref{ecdfloglog} [red for $z_i = 0$ and sky for $z_i = 1$]. The left hand side figure shows that $F_{0m} \geq F_n$ and the right hand side figure shows that the $\log{(-\log{F(x)})}$ functions differ by a constant. The scores produced by the classifier for the individuals in each group indicates the proportional reversed hazard model. Moreover, under the null hypothesis ($\theta=1$), the computed value of the test statistic $Z_w$ is $17.0741$ and the corresponding $p$-value ($\approx 0$) suggests the rejection of the null hypothesis.

\begin{figure}[ht]
    \centering
    \includegraphics[scale=.5]{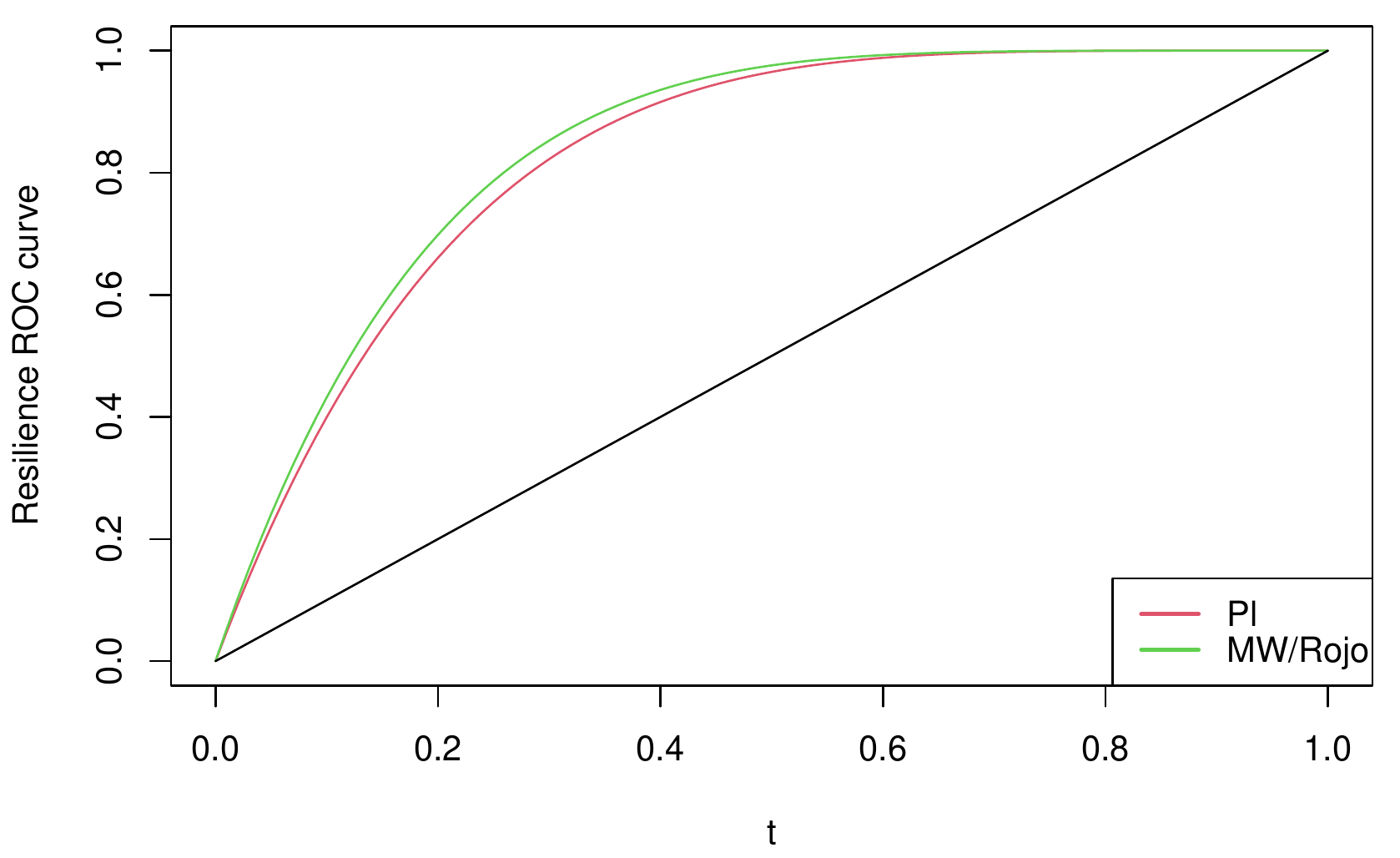}
    \caption{ROC plots based on Pl method and Mann-Whitney statistics}
    \label{plmw}
\end{figure}

\begin{table}[ht!]
\centering
\caption{Summary index}
\label{IDMTTFtable}
\begin{tabular}{clcccc} 
\hline
\begin{tabular}[c]{@{}c@{}}\\\\Method\end{tabular} &  & $\hat{\theta}$ (confidence interval) & AUC ($\hat{\tau}$) & Youden index ($\hat{J}$)~ & \multicolumn{1}{l}{Youden index point}  \\ 
\hline
Pl                                                 &  & 4.8444 (4.2883,~5.4005)              & 0.8289             & 0.5264                    & (0.3366, 0.8631)                        \\
MW                                                 &  & 5.3707 (4.3091, 6.4322)              & 0.8430             & 0.5540                    & (0.3193, 0.8733)                        \\
Rojo                                               &  & 5.3707 (4.3091, 6.4322)              & 0.8430             & 0.5540                    & (0.3193, 0.8733)                        \\
\hline
\end{tabular}
\end{table}

Table \ref{IDMTTFtable} presents all the estimated summary index of the ROC curve including resilience parameter $\theta$. Table \ref{IDMTTFtable} shows that all the estimates based on MW statistic and Rojo approach are the same. Figure \ref{plmw} shows the ROC plots based on partial likelihood and Mann-Whitney statistics.

\begin{figure}[ht]
    \centering
    \includegraphics[scale=.62]{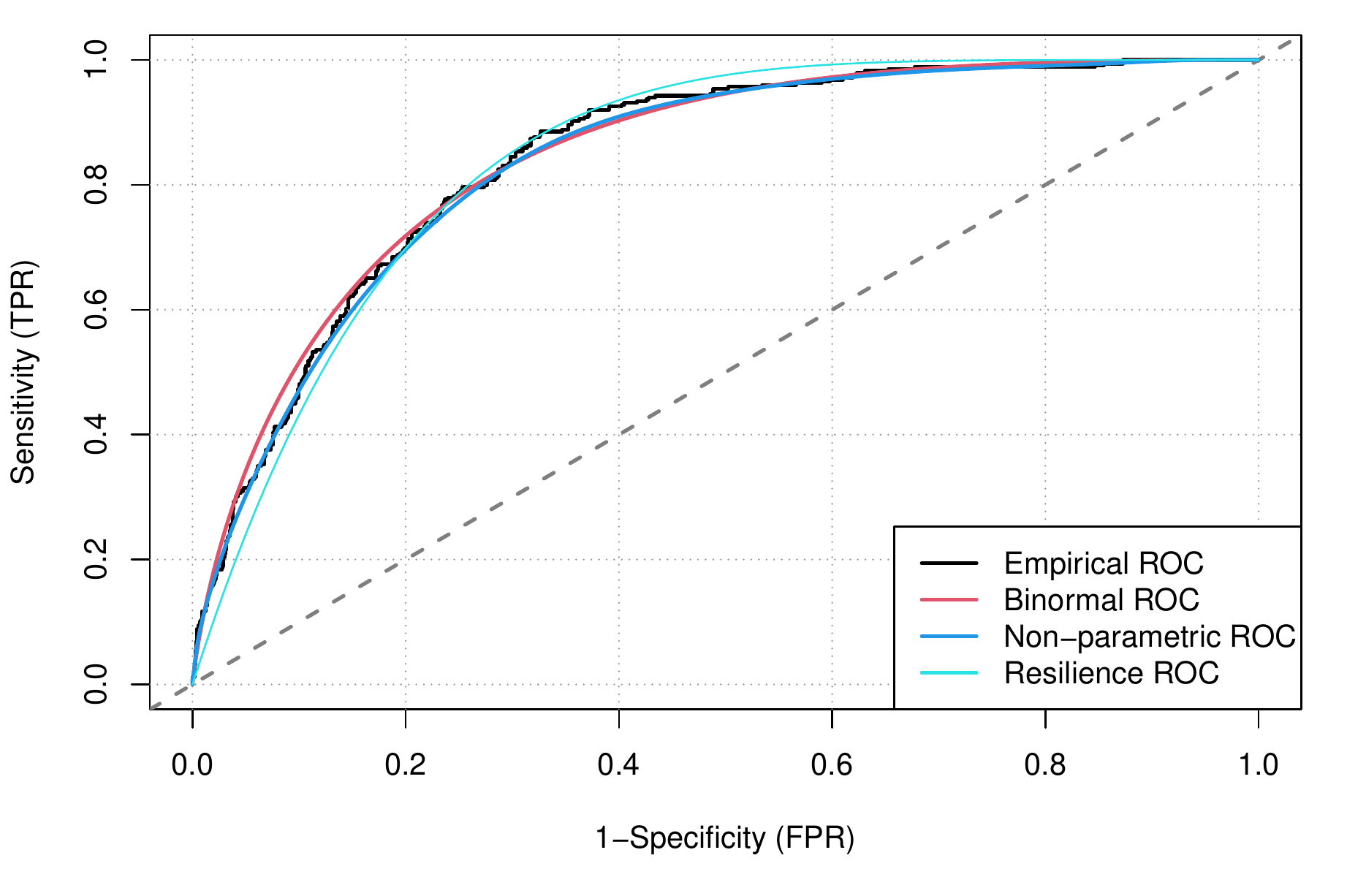}
    \caption{Plot of ROC curves based on existing methods and the proposed model}
    \label{allroc}
\end{figure}

\begin{figure}[ht!]
    \centering
    \includegraphics[scale=.62]{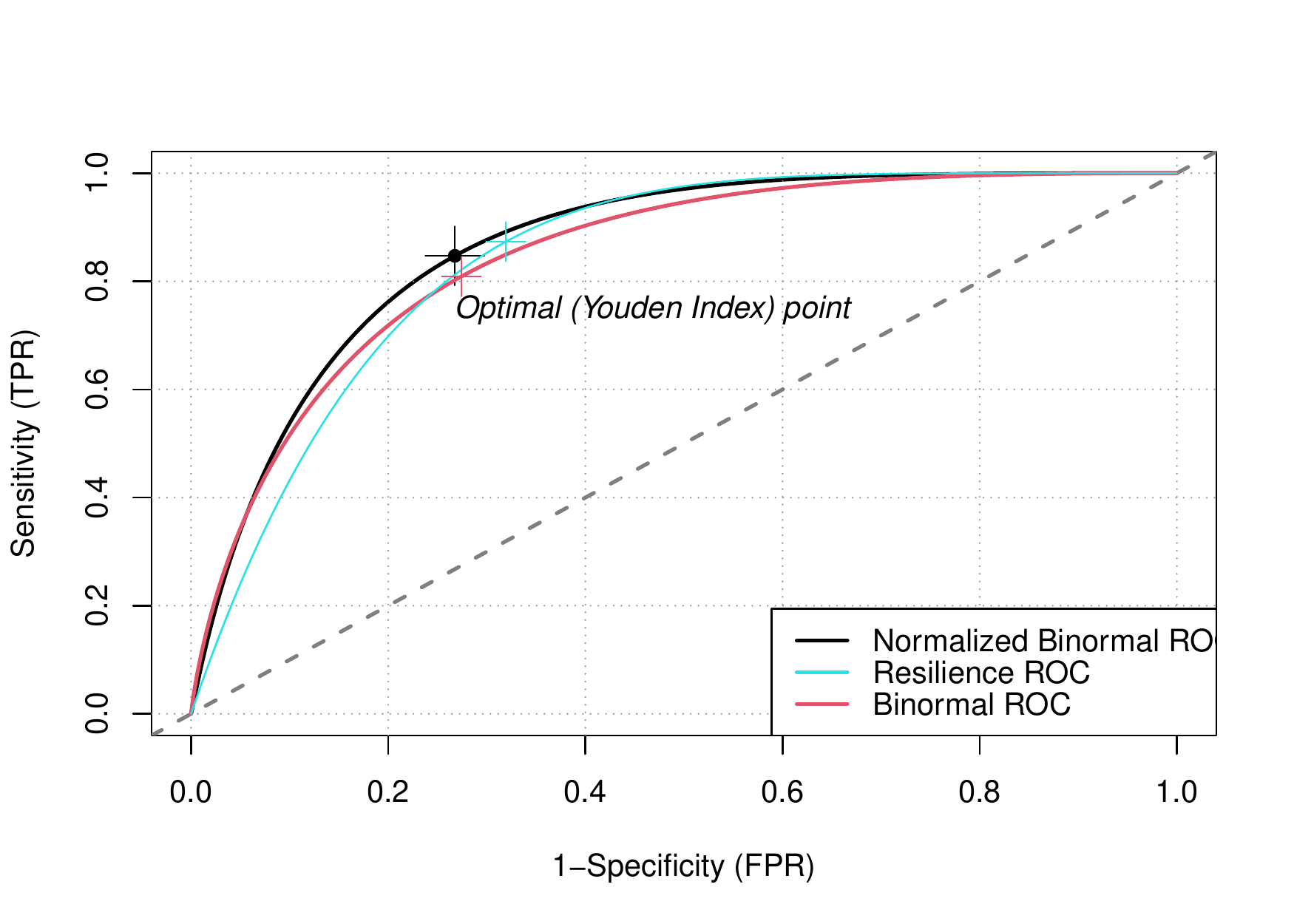}
    \caption{Normalized binormal ROC, resilience ROC and binormal ROC curve with Youden index point}
    \label{ResiBi}
\end{figure}

We have used `ROCit' package to estimate the values of the area under curve (AUC) based on existing method (\cite{rit}). The estimated values are 0.843, 0.8455 and 0.8375 using empirical, binormal and non-parametric methods. Figure \ref{allroc} represents all the ROC curves evaluated  based on empirical method, binormal assumption, non-parametric method and the proposed model. Note that in this figure the estimate of $\theta$ is chosen based on Mann-Whitney statistics. Figure \ref{allroc} depicts that the proposed model performs quite impressively in spite of having a very simple analytic form.

The invariant property of ROC curve under an increasing transformation motivates us to normalize the data for obtaining better results. Thus, in order to make the data into a normal shape the Yeo-Johnson transformation has been used since it does not require the input data to be positive unlike Box-Cox transformation (see \cite{yeo2000new}). For this purpose, the recommended package in R is `bestNormalize'  (\cite{Ryan}). %lambda = 0.8861159 (TPR0), lambda = 0.973429 (FPR1) 

Figure \ref{ResiBi} shows normalized  binormal ROC, resilience ROC and binormal ROC curve with corresponding Youden index point. Here `normalized  Binormal ROC' and `Binormal ROC' imply the binormal ROC curve with normalized data and without normalized data respectively. Moreover, this figure contains the resilience ROC curve based on Mann–Whitney statistics or Rojo approach since the AUC is more closer to the other existing models. Note that the Figure \ref{ResiBi} indicates that the resilience ROC curve performs similarly in comparison with the binormal model.

%For binormality assumption, one can also use Yeo-Johnson Transformation in order to make the data into a normal shape (see \cite{yeo2000new}). For this purpose we recommend the `bestNormalize' package in R (\cite{Ryan}). We have not used this transformation since the data exhibits normality and does not change AUC and the ROC curve significantly.

\subsection{Data in the context of  brain injury-related biomarkers}
The main focus of this subsection is to compare the proposed model with the Lehmann family of the ROC curves. At this juncture, it is reasonable to consider the data of \cite{turck2010multiparameter} since this data has been analysed by \cite{jokiel2020estimation} for the Lehmann family of the ROC curves. \cite{turck2010multiparameter} conducted a study for outcome prediction following aneurysmal subarachnoid hemorrhage (aSAH) using a combination of clinical scores together with brain injury-related biomarkers 113 patients admitted within 48 hours. After six months, based on the the condition of patients, the outcome was categorised as good when the Glasgow Outcome Scale
(GOS) is greater than equal to $4$ (41 observations) or poor (72 observations) otherwise. This data can be found in pROC package (\cite{robin2011proc}). To illustrate the proposed method, I will also consider nucleoside diphosphate kinase A (NDKA) level as the marker. \cite{jokiel2020estimation} pointed out that hypotheses of normality of the NDKA level distribution in any of two groups were rejected. %The following graph shows the log-log plot of empirical survival and empirical distribution function.

\begin{figure}[ht]
  \begin{minipage}{0.45\textwidth}
    \includegraphics[scale=.45]{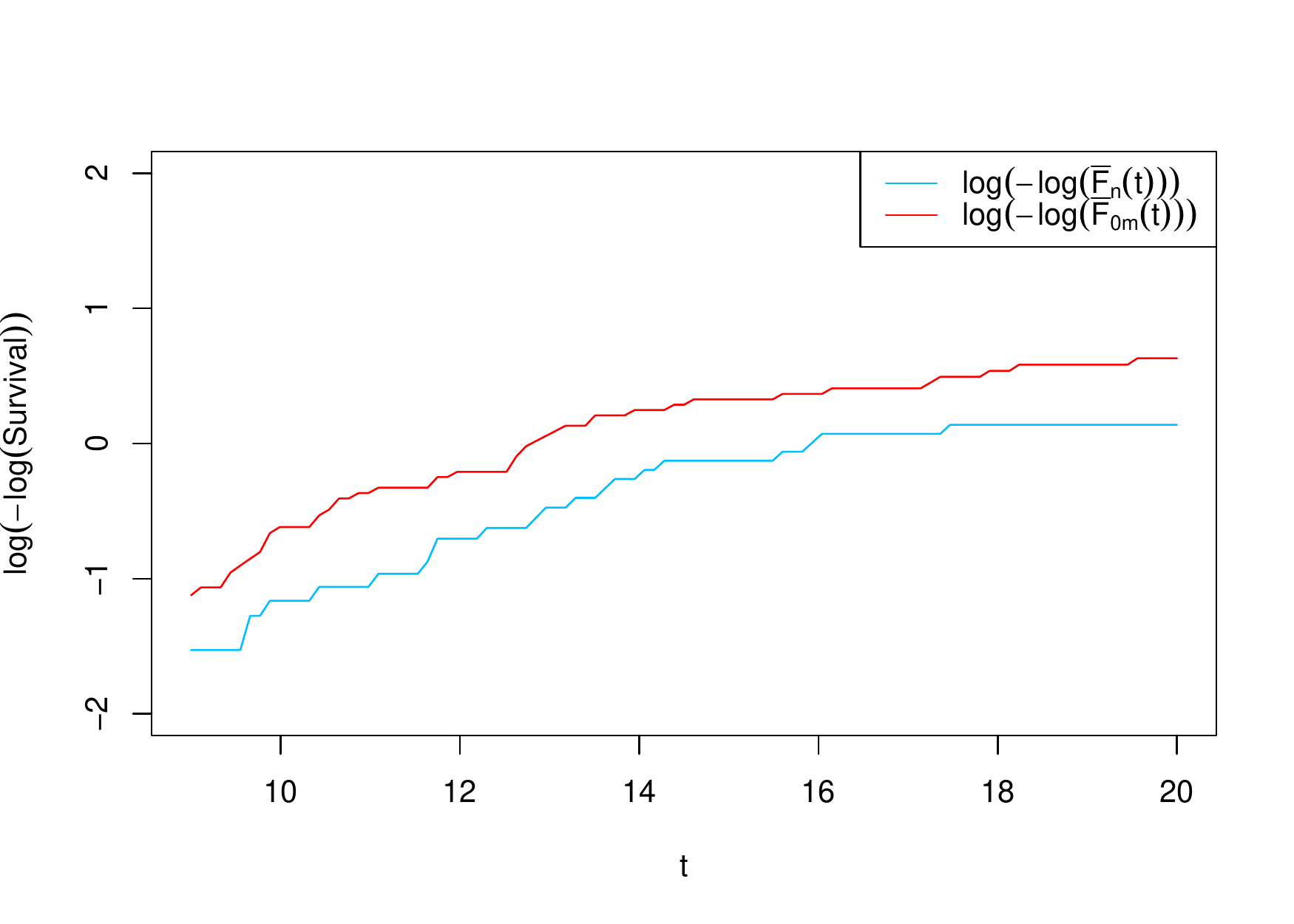}
    \label{fig11}
  \end{minipage}
  \hspace{.5cm}
  \begin{minipage}{0.45\textwidth}
    \includegraphics[scale=.45]{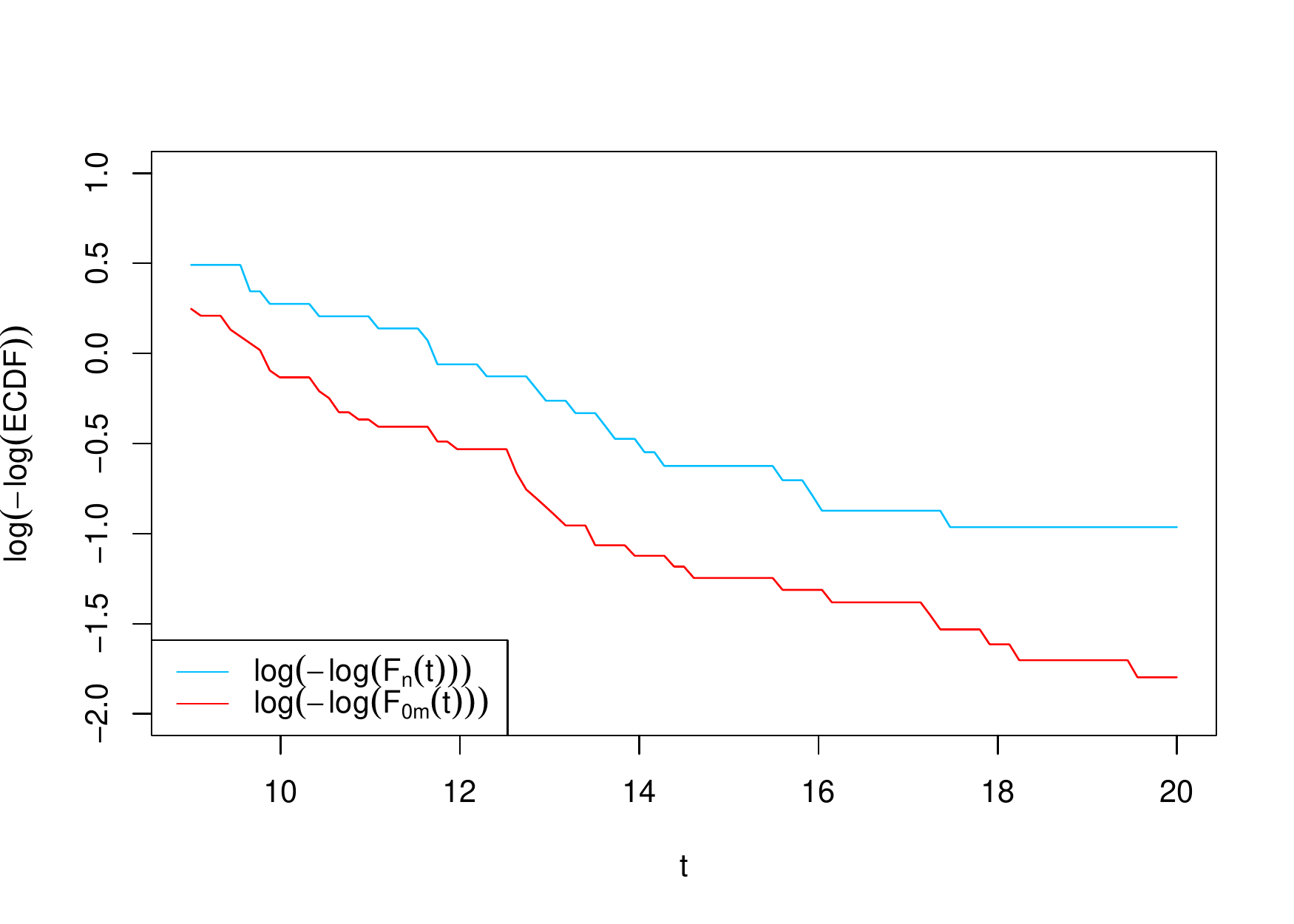}
    \label{fig22}
  \end{minipage}
      \caption{log-log plot of empirical survival and empirical distribution functions}
      \label{log_turck}
\end{figure}

\begin{figure}[H]
    \centering
    \includegraphics[scale=.48]{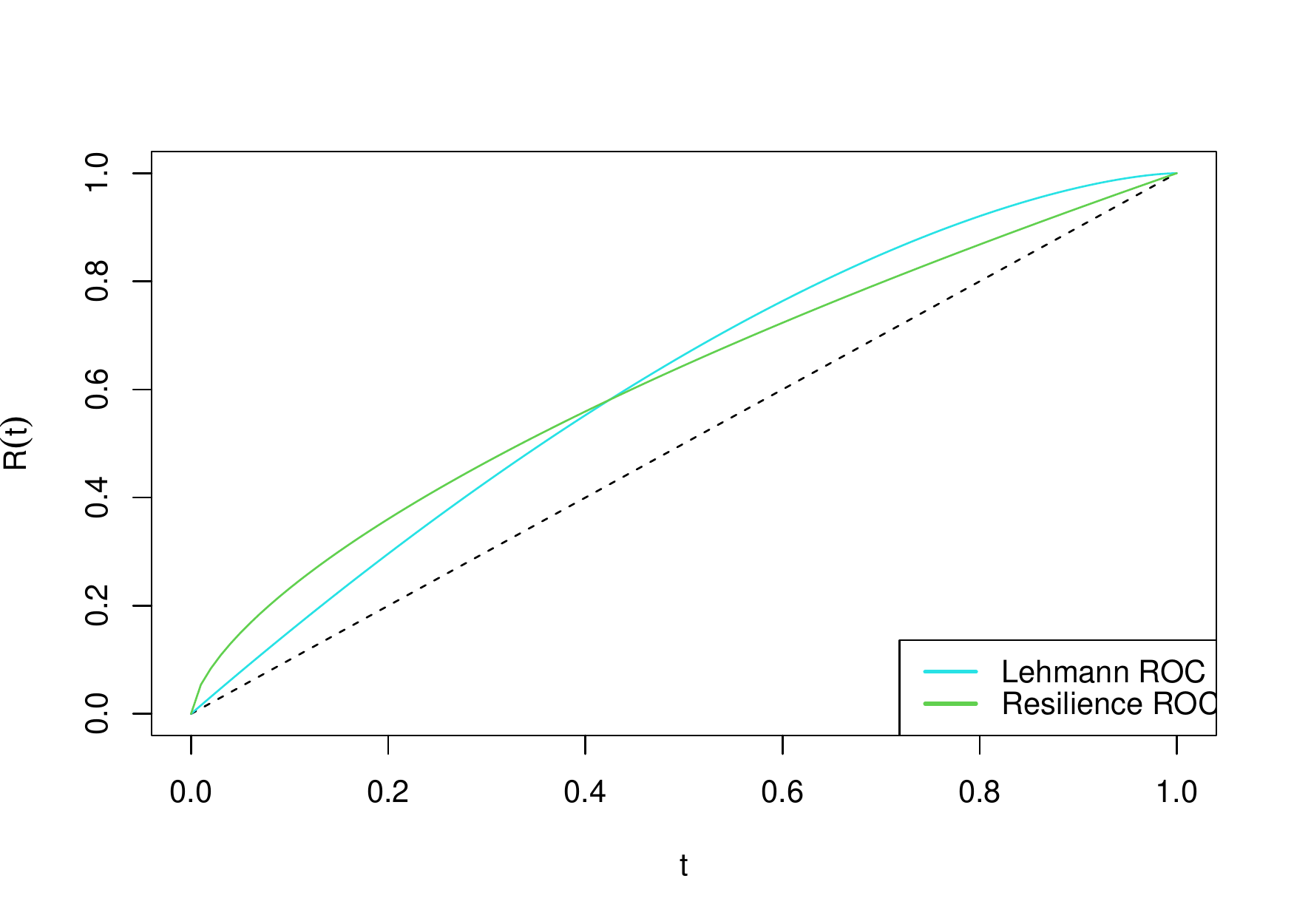}
    \caption{ROC plots based on Mann-Whitney statistics for Lehmann and Resilience model}
    \label{rocturck}
\end{figure}

Figure \ref{log_turck} indicates that both the groups support the proportional hazard and proportional reversed hazard assumption. Table \ref{summaryturck} shows all the estimated summary indices of the resilience ROC curve. Note that the estimated values are quite closer for MW and Rojo method while the lowest estimate was observed for partial likelihood method. Figure \ref{rocturck} displays estimated Lehmann ROC curve and resilience ROC curve based on MW method. Moreover, this figure depicts that the Lehmann ROC curve dominates resilience ROC curve before $t=0.4244$ and the resilience ROC curve dominates the Lehmann ROC curve $t=0.4244$. We can conclude that the measurement of NDKA level has strong evidence in order to predict aSAH.

\begin{table}[ht]
\centering
\caption{Summary index}
\label{summaryturck}
\begin{tabular}{clcccc} 
\hline
\begin{tabular}[c]{@{}c@{}}\\\\Method\end{tabular} &  & $\hat{\theta}$ (confidence interval) & AUC ($\hat{\tau}$) & Youden index ($\hat{J}$)~ & \multicolumn{1}{l}{Youden index point}  \\ 
\hline
Pl  &  & 1.3000 (0.7964,~1.8036)     & 0.5652    & 0.0962   & (0.3366, 0.8631)                        \\
MW  &  & 1.5737 (0.8483, 2.2991)  & 0.6114 & 0.1654  & (0.4170, 0.3208)                        \\
Rojo &  & 1.5826 (0.8526, 2.3127) &  0.6128 & 0.1674  & (0.2874, 0.4548)                        \\
\hline
\end{tabular}
\end{table}

\subsection{Ducheme muscular dystrophy data}
\begin{figure}[H]
    \centering
    \includegraphics[scale=.55]{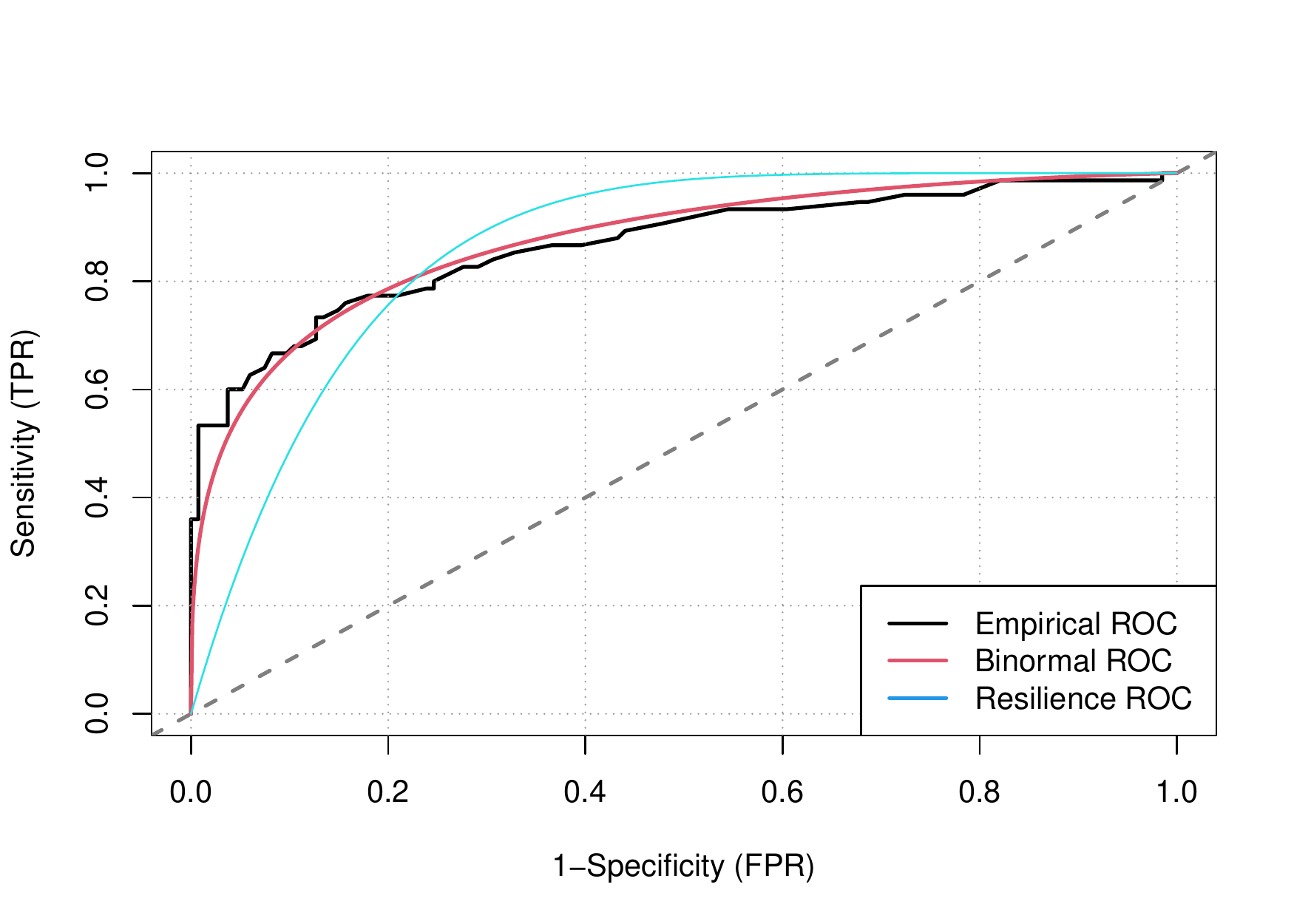}
    \caption{Plot of standardised binormal ROC curve, empirical ROC curve and proposed resilience ROC curve based on Mann-Whitney statistics}
    \label{rocck}
\end{figure}

In this subsection, our aim is to address a misspecification scenario. Ducheme muscular dystrophy (DMD) disease causes rapid progression of muscle degeneration of a child. It is a genetically transmitted disease from a mother to her child. It is a well-known muscular dystrophy and there is no cure at all for this disease. Thus, it is very important to diagnose all affected females. \cite{andrews2012data} reported DMD data in Table 38.1 which was collected during a program conducted at a hospital for sick children in Toronto. In our study, we consider one of the serum enzyme levels, i.e., creatine kinase (CK), for 75 carriers and 134 noncarriers (healthy females) as a bio-marker. The following Figure \ref{rocckprhr} indicates non-proportionality in the reversed hazard rate. Consequently, we are interested to apply the proposed model under misspecification together with existing models. Here we consider binormal ROC curve and empirical ROC curve for comparison purpose. Moreover, in order to make the data into a normal shape the Yeo-Johnson transformation has been used for binormal ROC curve (see \cite{yeo2000new}). The AUCs for binormal ROC curve and empirical ROC curve are 0.8741 and 0.8679 respectively. All the estimated summary indices of the resilience ROC curves are reported in Table \ref{cksum}. 
Table \ref{cksum} shows that all the estimates based on MW statistic and Rojo approach are same as expected. The Figure \ref{rocck} indicates that based on MW statistic and Rojo approach the resilience ROC curve are also able to handle misspecification scenario.
\begin{figure}[H]
    \centering
    \includegraphics[scale=.55]{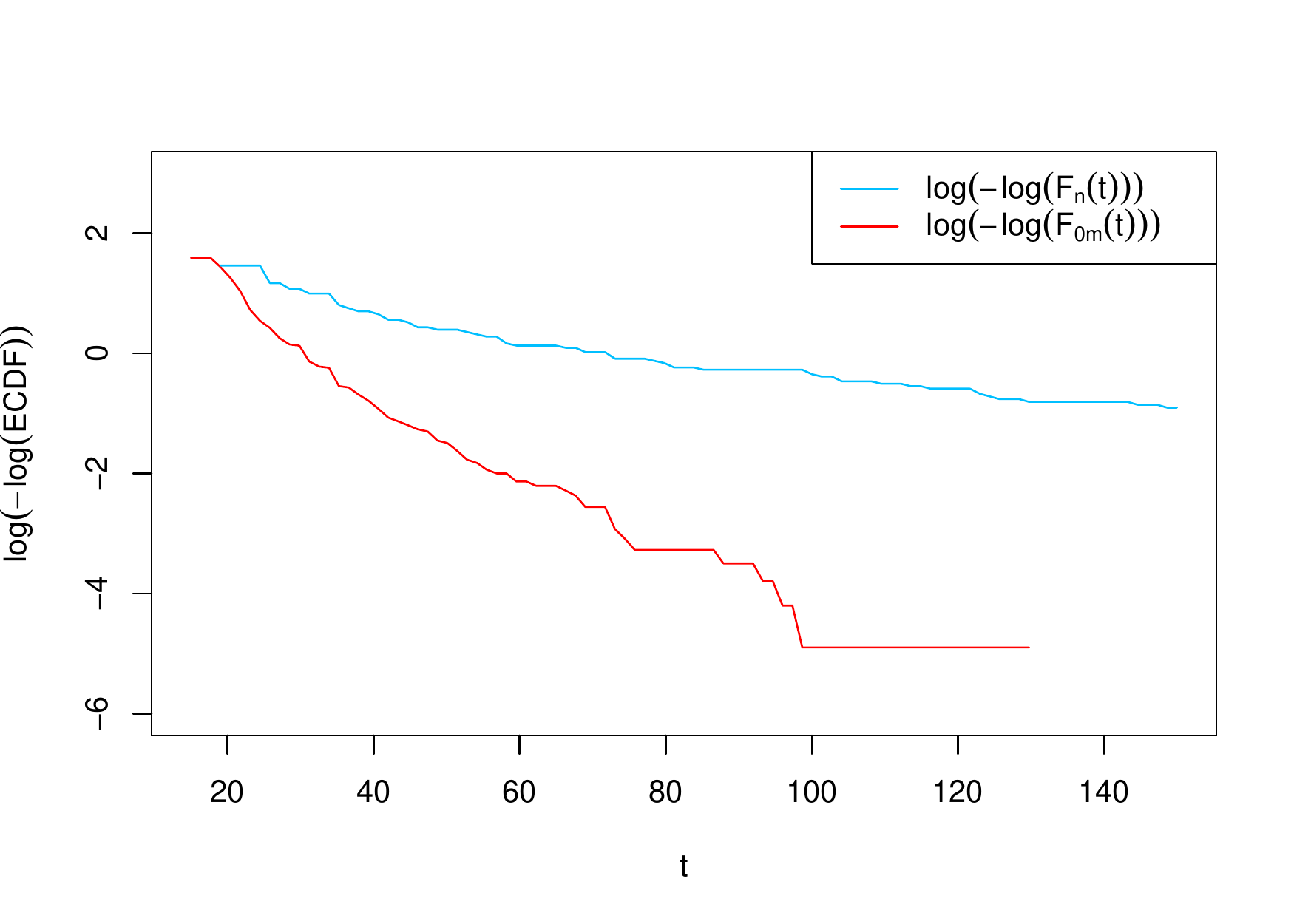}
    \caption{log-log plot of empirical distribution functions}
    \label{rocckprhr}
\end{figure}

\begin{table}[ht]
\centering
\caption{Summary index}
\label{cksum}
\begin{tabular}{clcccc} 
\hline
\begin{tabular}[c]{@{}c@{}}\\Method\end{tabular} &  & $\hat{\theta}$ (confidence interval) & AUC ($\hat{\tau}$) & Youden index ($\hat{J}$)~ & \multicolumn{1}{l}{Youden index point}  \\ 
\hline
Pl                                                   &  & 3.4702 (4.4973, 2.4431)              &  0.7763                  & 0.4302                           &  (0.1741, 0.6043)                                       \\
MW                                                   &  & 6.3304 (9.3709, 3.2899)              & 0.8636             &   0.5956                        &   (0.1117, 0.7074)                                      \\
Rojo                                                 &  & 6.3304 (9.3709, 3.2899)              & 0.8636             &  0.5956                          &   (0.1117, 0.7074)                                       \\
\hline
\end{tabular}
\end{table}

\section{Discussion}

The paper proposes a new semiparametric model of the ROC curve which is an alternative to the existing models based on the resilience parameter family  or alternatively, a proportional reverse hazards family, with underlying distribution $F_0$. The proposed model does not require a full
parametric specification of the distribution of the scores produced by the binary classifier for the two reference populations. The resulting ROC curve and its summary indices (such as AUC and Youden index) have simple analytic forms. A brief discussion about the verification of the PRHR assumption has been discussed. The partial likelihood method is applied to estimate the ROC curve. This method has some advantages for covariate adjustment due to its regression representation since it is possible to formulate most practical ROC problems using a regression model. The estimation methodology has been developed for the AUC exploiting Mann-Whitney statistics and the Rojo approach. Estimation procedure of the Youden index is also presented based on the proposed estimation methods. A simulation study has been conducted to assess the performance of all the considered estimators. Moreover, three real data sets have been analysed based on the proposed model and existing models together with some important implications therein. 

Now the discussion about the practical importance of the choice between binormal and resilience family of ROC curve is crucial. The resilience family of ROC curve can be applicable if the scores produced by the classifier for the individuals in each group indicates the proportional reversed hazard model regardless of the binormality assumption. Moreover, the resilience family of ROC curve will be most effective when the hypotheses of normality in any of two groups were rejected based on Shapiro-Wilk test and the two population groups hold the PRHR assumption. In the proposed model, the expression of the ROC  curve  has very simple analytic form and all the summary indices can be calculated quickly. Thus the resilience family of ROC curve can be regarded as a potential tool in ROC curve analysis.

This paper proposes ROC curve analysis in the direction of resilience family of distributions for the first time. A lot of work needs to be carried out as a further research for the proposed model. %In this paper, the PRHR assumption is verified graphically but development of a test for the verification of the PRHR assumption is necessary to undertake as a future work. 
Development of the Bayesian estimation methodology for the resilience parameter (assuming $\theta\in (1, \infty$)) could be considered. The problem of estimation of $\theta$ for the resilience family of the ROC curves based on smoothed empirical distribution function or minimum distance method using the technique similar to that in  works of \cite{jokiel2013nonparametric} and \cite{jokiel2021minimum} is also of great interest. Moreover, it is well worth to extend the estimation methodologies for the resilience ROC curves under left censored data since \cite{kalbfleisch1989inference} established that the property of RHR is important in the estimation of the survival function under left censored data. 

\section*{Acknowledgements}
 The present author is grateful to the Theoretical Statistics and Mathematics Unit of Indian Statistical Institute, Delhi Centre, India for offering a visiting position and providing necessary infrastructure.

\bibliographystyle{apa}
\bibliography{winnower_template}
\appendix

\end{document}